\documentclass[a4paper,12pt]{article}
\DeclareMathAlphabet{\Bbb}{U}{msb}{m}{n}
\DeclareMathAlphabet{\euf}{U}{euf}{b}{n}
\DeclareMathAlphabet{\bi}{OML}{cmm}{b}{it}
\makeatletter
\def\^{{}^}
\def\_{{}_}
\def\alg#1({\euf{#1}(}
\def\grp#1({\mathsf{#1}(}
\def\ft#1#2{{\textstyle{{#1}\over{#2}}}}
\def\del{\partial}
\def\cc{^*\!\!}
\def\d{\mathrm{d}}
\def\i{\mathrm{i}}
\def\e#1#2{e_{#1}\^{#2}}
\def\ee#1#2{\tilde e_{#1}\^{#2}}
\def\ce#1#2{\tilde c_{#1}\^{#2}}
\def\re#1#2{\tilde r_{#1}\^{#2}}
\def\E#1#2{E_{#1}\^{#2}}
\def\EE#1#2{\tilde E_{#1}\^{#2}}
\def\GG{\tilde G}
\def\gg{\tilde g}
\def\O{\mathnormal{\Omega}}
\def\rot{\lambda}
\def\irot{\sigma}
\def\Rot{\mathnormal{\Lambda}}
\def\V{\xi}
\def\W{\zeta}
\def\eps{\varepsilon}
\def\RR{\Bbb{R}}
\def\CC{\Bbb{C}}
\def\MM{\Bbb{M}}
\def\fut{\mathcal{F}}
\def\t{\theta}
\def\Re{\mathrm{Re}\,}

\def\M{\mathcal{M}}
\def\N{\mathcal{N}}
\def\X{\bi{x}}
\def\diag{\mathrm{diag}}

\def\lag{\mathcal{L}}
\def\Lmat{I_{\mathrm{mat}}}
\@addtoreset{equation}{section}
\def\theequation{\arabic{section}.\arabic{equation}}
\def\gllabel[#1]{\label{#1}}
\def\beq{\arraycolsep .1em\begin{eqnarray}\@ifnextchar[{\gllabel}{}}
\def\eeq{\end{eqnarray}}
\def\zl{\nonumber\\}
\def\txt#1{\quad\hbox{#1}\quad}
\def\eref#1{(\ref{#1})}
\def\sref#1{section~\ref{#1}}
\def\ssty{\scriptstyle}
\makeatother

\begin{document}
\begin{flushright}
KCL-TH-95-10\\
gr-qc/9511034
\end{flushright}
\vfill
{\LARGE \bf
\begin{center}
Causal~structure~and~diffeomorphisms in~Ashtekar's~gravity\\[2ex]
\end{center}}
\begin{center}
        Hans-J\"urgen Matschull\\[1ex]
        Department of Mathematics,
	King's College London,\\
	Strand,
	London WC2R 2LS,
	England\\[1ex]
        November 9th, 1995\\[4ex]
\end{center}
\begin{abstract}
A manifestly diffeomorphism invariant extension of Einstein gravity is
constructed, which includes singular metrics, and whose ADM formulation is
Ashtekar's gravity. The latter is shown to be locally equivalent to the
covariant theory. It turns out that exactly those kinds of degenerate four
dimensional metrics are allowed which do not destroy the causal structure of
spacetime. It is also shown that Ashtekar's gravity possesses an extension that
provides a local $\grp SO(3,\CC)$ invariance, without complexifying or changing
the signature of the metric.
\end{abstract}

\vfill

\section{What is the problem?}\label{one}
The purpose of this paper is to reexamine a question considered by
Bengtsson~\cite{bengtsson:89b} some time ago, which also appears in the recent
work of Reisenberger~\cite{reisenberger:95}, and which has some interesting
relations to the ideas of Dragon~\cite{dragon:92}. The question is whether
there is a manifestly diffeomorphism invariant formulation of Ashtekar's
`polynomial' gravity, where the metric is real Lorentzian but possibly
degenerate. More precisely, the question is whether Ashtekar's gravity is the
`ADM'~\cite{arnowitt.deser.misner:62} formulation of some covariant theory,
obtained in the same way as the usual ADM formulation of gravity is obtained
from Einstein's theory. Bengtsson investigated this question and came to the
conclusion that a theory exists which is equivalent to Ashtekar's theory for
complexified gravity and non-vanishing lapse function. Reisenberger has shown
that (even without the last restriction) Ashtekar's theory is invariant under
{\em infinitesimal} diffeomorphisms, but he pointed out some problems
concerning finite diffeomorphisms. He argued that these problems are related to
the causal structure of spacetime. Finally, Dragon considered, quite generally,
manifestly covariant extensions of Einstein gravity allowing degenerate
metrics. In his results, the close relation to Ashtekar's variables is not
apparent. Here I will show that there is an improved version of Bengtsson's
theory, which is an extension of {\em real\/} Einstein gravity, and where the
restriction `lapse${}\neq0$' is replaced by a fully covariant condition, which
implies that spacetime has a well defined causal structure, at least locally.
The resulting theory can equivalently be considered as a restricted version of
the vierbein formulation of Dragon's theory.

\subsection*{Ashtekar's theory}
To set up the notation, let me briefly review what the problem actually is.
As is well known, the Lagrange density of Ashtekar's gravity can be written
as~\cite{ashtekar}
\beq[lag]
   \lag = \i \ee am \, \del_t A_{ma}
         +  \i A_{ta} \,   D_m \ee am
         -  \i N^m \, \ee an F_{mna}
         + \ft12 N \, \eps_{abc}\, \ee am \ee bn \, F_{mnc},
\eeq
where spacetime is split into a spatial hypersurface $\N$ with local
coordinates labeled by indices $m,n,\dots$ and a time coordinate $t$. The basic
fields are the $\alg so(3,\CC)$ connection $A_{\mu a}$ (where $\mu,\nu,\dots$
are spacetime indices taking the `values' $m,n,\dots$ and $t$), the real
densitized inverse dreibein $\ee am=e \e am$ (where $e$ is the determinant of
the spatial dreibein $e_{ma}$) and the lapse and shift fields $N$ and $N^m$. As
the lapse function represents a pure gauge degree of freedom, we can restrict
it by $N>0$. As mentioned above, $N\neq0$ is a restriction already found
in~\cite{bengtsson:89b}, for the transformation to the covariant theory to
exist. The sign fixing does not impose any additional constraint, as long as
$N$ is continuous.

The flat indices $a,b,\dots$ take the values $1,2,3$, and $\eps_{abc}$ are the
structure constants of $\alg so(3)$, defining the covariant derivative and
field strength
\beq[field-strength]
    D_m \ee am &=& \del_m \ee am + \eps_{abc} A_{mb} \ee cm, \zl
  F_{mna}& = &\del_m A_{na} - \del_n A_{ma} + \eps_{abc} A_{mb} A_{nc}.
\eeq
In the following, I will call this theory  `Ashtekar's gravity'.
Its maybe most interesting feature is the polynomial form of the action  in
terms of the canonical variables, provided that $t$ it used as the canonical
time variable. This makes it possible to take it as an extension of Einstein
gravity, allowing certain kinds of singular metrics, corresponding to
non-invertible matrices $\ee am$. Thereby the densitized spatial metric
$\gg^{mn}=\ee am \ee an$ becomes degenerate as well, but remains positive
semidefinite. Hence, for canonical formulation and quantization of gravity, the
action \eref{lag} has turned out to be a promising starting point. Here I want
to proceed into the opposite direction. The main question to be considered is
whether there is a {\em manifestly\/} diffeomorphism invariant theory which,
after introducing a spacetime slicing `a la ADM', leads to the Lagrangian
\eref{lag}. Having such a theory, we do not only get an elegant proof for the
diffeomorphism invariance of Ashtekar's gravity, which will be a simple
consequence of the manifest invariance of the `higher' theory. It will also
provide a deeper understanding of its geometrical structure, and it is this I
want to focus on. In particular, we will see what the singular metrics `look
like' in four dimensions and how they affect the causal structure of spacetime.

\subsection*{Diffeomorphisms}
The first important question is whether Ashtekar's theory is really invariant
under diffeomorphisms of spacetime. This is not obvious because to write down
the action one has to introduce a {\em particular\/} coordinate $t$ as a
background structure. You can think of it as given by a scalar field $T(\X)$ on
spacetime, subject to the condition $\del_\mu T\neq0$, which provides the ADM
slices as the $T={}$constant hypersurfaces. This time coordinate not only
appears explicitly in the action. It is also required to define the fields
themselves. Whereas the connection can be combined into a four dimensional
one-form, there is no covariant object which is linear in, or even a
homogeneous polynomial of the dreibein and the lapse and shift fields. However,
we know that the action reduces to the Einstein Hilbert action for invertible
dreibeins (and after solving the equation of motion for $A_{\mu a}$), so in
this case it {\em is\/} invariant under the full four dimensional
diffeomorphism group. In addition, the set of invertible dreibeins is dense
inside the set of all dreibeins, and the action is obviously continuous in $\ee
am$ (and even analytic). Hence, we conclude that the theory must be invariant,
provided that the symmetry transformations themselves are well behaving on the
`boundary', i.e.\ when the dreibein becomes singular. It is at this point where
the actual problem arises: because the field content of the theory itself
depends on the chosen background $T(\X)$, it is not obvious that a field
configuration (with singular metric) defined with respect to one slicing has a
representation in another slicing.

A straightforward way to find out whether the transformation under
diffeomorphisms is  continuous in $\ee am$, $N$, and $N^m$, is to compute the
transformation explicitly. Let $\W^\mu=(\W^t,\W^m)$ be a generating vector
field of an infinitesimal transformation. Then $A_{\mu a}$ should transform by
its four dimensional Lie derivative. If one takes this as an ansatz and
computes the transformation of the dreibein, one finds that it becomes complex.
This is because the Lagrangian \eref{lag} involves some kind of gauge fixing,
which will be discussed in detail below. However, one can compensate for the
imaginary contributions by introducing an additional $\grp SO(3,\CC)$ rotation,
with field dependent parameter. As a consequence, all the transformations
become highly non-linear, but polynomial\footnote{Unfortunately, a
contradictory result obtained in a previous paper~\cite{matschull:94a} has
turned out to be wrong. The correct result has been derived
in~\cite{reisenberger:95}, which focusses on gauge transformations generated by
the canonical constraints instead of directly computing field transformations
induced by four dimensional diffeomorphisms. This leads to different formulae
for the canonical fields (which coincide with \eref{infi} if the equations of
motion are satisfied) but identical formulae for the multipliers.}. The
explicit formulae are
\beq[infi]
\delta N &=& \W^n \, \del_n N - \del_n \W^n \, N
            + \del_t (\W^t N)
            - 2 \, \del_n \W^t \, N N^n, \zl
\delta N^m &=& \W^n \, \del_n N^m - \del_n \W^m \, N^n
            + \del_t (\W^t N^m) \zl && \hspace{7em}{}
            - \del_n \W^t \, ( N^2 \gg^{mn} + N^m N^n ) + \del_t \W^m , \zl
\delta \ee am &=& \del_n (\W^n \ee am) - \del_n \W^m \, \ee an
            + \W^t \, \del_t \ee am \zl && \hspace{7em}{}
            + \del_n \W^t \, ( \ee am N^n - \ee an N^m ),
\eeq
for the metric fields. In each expression, the first two terms are the
transformations of the fields under spatial diffeomorphisms. The third terms
describe the behaviour under pure rescaling of the time coordinate: the lapse
and shift functions are densities of weight one with respect to these
transformations, whereas $\ee am$ has weight zero. The last term in $\delta
N^m$ can be understood by interpreting $N^m$ as the gauge field associated with
spatial diffeomorphisms. The remaining nonlinear terms appear whenever
$\del_n\W^t\neq0$. This means that the time rescaling depends on the space
point, so that the slicing itself is affected by the diffeomorphism. A similar
term appears in the transformation of the connection, which explicitly shows
the compensating gauge transformation:
\beq
  \delta A_{\mu a} = \W^\nu \del_\nu A_{\mu a}
                    + \del_\mu \W^\nu A_{\nu a}
              + D_\mu (\i \del_n \W^t N \ee an ).
\eeq
Without the last term, we had to add the corresponding rotation to $\delta\ee
am$, which would produce an imaginary contribution. A cumbersome but
straightforward computation now shows that the action is in fact invariant (up
to a total time derivative) under these transformations. Hence, we have well
defined infinitesimal diffeomorphisms, but because of the higher order terms it
is not guaranteed that they can be integrated to give finite diffeomorphisms.
Ashtekar's theory is therefore invariant at least under `small'
diffeomorphisms, but from the somewhat awkward formulae \eref{infi} we do not
learn much about the geometrical nature of the theory, especially it is hardly
possible to see what `small' means, i.e.\ whether \eref{infi} can be integrated
or not for a given generating vector field $\W^\mu$.

In fact, for very special solutions of the field equations, it has been shown
in~\cite{reisenberger:95} that there are finite diffeomorphisms connected to
the identity for which \eref{infi} can {\em not\/} be integrated. It is shown
that this is similar to a well known feature of the ADM formulation of metric
gravity. There, the slices have to be spacelike and a given solution in one
slicing cannot be transformed into another slicing unless the new slices are
everywhere spacelike. In addition, the pure existence of {\em some\/} slicing
already restricts the possible filed configurations: spacetime must be time
orientable and closed causal (i.e.\ positive oriented timelike or lightlike)
curves are forbidden. By constructing the fully covariant theory associated
with Ashtekar's gravity below, we will see that it is {\em exactly\/} this what
happens here, too. There will be a well defined `causal structure' on
spacetime, and the restrictions we get are identical to those of ADM gravity.

Instead of using infinite diffeomorphisms and integrating them, it might be
more convenient to consider finite diffeomorphisms right from the beginning.
Moreover, instead of computing the transformations of the fields under
diffeomorphisms explicitly, it is more suitable to construct covariant objects,
i.e.\ proper four dimensional tensors, whose transformations laws then become
very simple. Let us see whether there is a tensor that carries the $\grp SO(3)$
invariant information about the dreibein, the lapse and the shift fields.
Hence, it should depend on $\gg^{mn}$, $N^m$ and $N$. These are $6+3+1=10$
independent components, so it is not surprising that the tensor is symmetric of
rank two~\cite{bengtsson:89b}:
\beq[G-def]
   \GG^{\mu\nu} = \pmatrix{ \GG^{tt} & \GG^{tn} \cr
                            \GG^{mt} & \GG^{mn} }
        = \pmatrix{ - N^{-1} &  N^{-1} N^n   \cr
                      N^{-1} N^m  & N \gg^{mn} - N^{-1} N^m N^n  }.
\eeq
For invertible metrics, this is nothing but the densitized inverse four
dimensional metric $\sqrt{-G}G^{\mu\nu}$, which appears in~\cite{dragon:92} as
the basic field variable. Here, the restriction $N>0$ is essential to provide
the correct sign.

What is the range of this tensor? First of all, we obviously have $\GG^{tt}<0$.
Moreover, because of the positivity of $\gg^{mn}$, the signature of
$\GG^{\mu\nu}$ is $(-,+,+,+)$, corresponding to invertible metrics, or some (or
all) of the `$+$' signs may be replaced by 0, leading to degenerate metrics.
Under these conditions the transformation back to Ashtekar's variables is
unique up to $\grp SO(3)$ rotations. The restriction $\GG^{tt}<0$, however, is
still non-covariant, because it refers to the special coordinate $t$. But we
already got some hint to how a covariant version could look like. There must be
exactly one `$-$' in the signature of $\GG^{\mu\nu}$. Hence, a covector
$\V_\mu$ must exist such that $\V_\mu\V_\nu\GG^{\mu\nu}<0$, which in some sense
(and in the usual sense for invertible metrics) means that there is at least
one `timelike direction'. I will not go into more details at this point, but we
can see that this will have something to do with timelike curves and causality,
and a restriction like this will be the central point of the covariant theory.

To restore the $\grp SO(3)$ gauge freedom, one can go back from $\GG^{\mu\nu}$
to its square root, which is linear instead of quadratic in $\ee am$. There is
an obvious square root of \eref{G-def}:
\beq[E-def]
  \EE A\mu = \pmatrix { \EE 0t & \EE 0m \cr
                        \EE at & \EE am }
    = \pmatrix { \sqrt N^{-1} &  -\sqrt N^{-1} N^m  \cr
                  0 & -\sqrt N \ee am }.
\eeq
Here, $\ssty A$ is a flat Lorentz index taking the values $0,1,2,3$ (and hence
$a$ refers to a subset of these values), which is raised and lowered, and has
to be contracted with the Lorentz metric $\eta^{AB}=\diag(-1,1,1,1)$ to give
$\GG^{\mu\nu}$. The signs are chosen for later convenience. In fact, this is
the basic variable used in~\cite{bengtsson:89b} to define a covariant theory.
It is a vector density of weight $\ft12$, and I will simply call it the
vierbein. However, as defined in \eref{E-def}, the vierbein is not yet a
covariant object, as it is subject to the conditions $\EE at=0$ and $\EE 0t>0$.
They still refer to the coordinate $t$. While the first condition can be
treated as a gauge fixing, the second one is somehow more subtle. Moreover,
they are not independent and the first one is not just a {\em Lorentz\/} gauge
fixing:
the transformation to the upper triangular form \eref{E-def} is not always
possible, as it already requires $\EE At$ to be timelike, or equivalently
$\GG^{tt}<0$. Hence, the gauge fixing already affects the diffeomorphism group.

One way out of this dilemma is to consider \eref{E-def} as a gauge fixed
version of an arbitrary {\em complex\/} vierbein (or alternatively switch to
Euclidean gravity), thereby extending the gauge symmetry to $\grp
SO(1,3,\CC)=\grp SO(4,\CC)$ (or replacing it by $\grp SO(4,\RR)$). This is done
in~\cite{bengtsson:89b} and also assumed in~\cite{reisenberger:95}. The
transformation to the upper triangular form is then always possible by a pure
$\grp SO(4)$ rotation. From the physical point of view, however, this is rather
unsatisfactory, as it takes us away from real Lorentzian gravity, and moreover
the complex theory is even different from complexified Einstein gravity: the
equivalence of Ashtekar's and Einstein gravity (for invertible metrics) is
based on the fact that the Einstein Hilbert action is the {\em real part\/} of
the complex action \eref{lag}, which is holomorphic in all the complex fields,
so the field equations are the same for both. However, this equivalence only
holds if the antiself-dual part of the connection is the complex conjugate of
the self-dual part, and this relation is lost when going over to complex
gravity with gauge group $\grp SO(4,\CC)$.

So the question of how to construct covariant objects out of the three
dimensional fields naturally leads us to complexification. Moreover, it is not
only the lack of a manifestly covariant formulation of Ashtekar's gravity that
is somehow unsatisfactory. Another weak point of \eref{lag} is that, though one
of the fields is complex and acts as an $\alg so(3,\CC)$ valued connection,
there is no local $\grp SO(3,\CC)$ invariance of the action. That is because
the dreibein has to be real, so we only have an $\grp SO(3,\RR)$ gauge
invariance, and as is also well known this leads to various problems with
reality conditions to be imposed on the canonical variables. Obviously, this
can also be solved by complexifying the fields. But as just described,
straightforwardly complexifying everything takes us to complex and non-Einstein
gravity. Nevertheless complexification is not a bad idea, and before coming to
the covariant formulation in \sref{cov}, I will show in the following section
that there is a different extension of Ashtekar's gravity, which has gauge
group $\grp SO(3,\CC)$ but does not take us away from positive semidefinite
metrics.

\section{Complexifying without complexifying}
Instead of simply making $\ee am$ complex, a local $\grp SO(3,\CC)$ invariance
can be obtained in a slightly different way and without making the metric
itself complex. The resulting action is a bit more complicated than \eref{lag},
but it has some advantages when considering Ashtekar's gravity as the ADM
formulation of its covariant formulation, i.e.\ when we are going away from the
canonical formalism back to a manifestly covariant theory. First of all, the
connection field really becomes an $\grp SO(3,\CC)$ gauge field, without
extending the physical phase space of the theory, i.e.\ we only need to add
extra {\em gauge\/} degrees of freedom. Secondly, there will be no gauge fixing
necessary when going over from the full four dimensional theory in terms of the
vierbein to the ADM formulation in terms of dreibein, lapse and shift. This
sounds a bit mysterious first, but note that $\grp SO(3,\CC) \simeq \grp
SO(1,3)_+$ is nothing but the four dimensional Lorentz group. So we have the
same gauge degrees of freedom in the covariant theory and in its ADM
formulation. It is only the representation of the gauge group which is
different.

The prize we have to pay for this is that on some of the fields the gauge
symmetry will be realized non-linearly. Hence, it is not clear whether this
kind of complexification is really suitable for the canonical treatment or even
quantization of Ashtekar's gravity, I just want to present it to show that
there is a formulation of real Lorentzian gravity in terms of Ashtekar's
variables which admits an $\grp SO(3,\CC)$ invariance, and that it can be
obtained as the ADM formulation of a covariant theory without gauge fixing. So
if you don't mind gauge fixing you can skip this section, which is not
necessary to understand the remainder of the article.

The basic idea is quite simple: consider the real dreibein $\ee am$, and act on
it with an arbitrary $\grp SO(3,\CC)$ transformation $\rot_{ab}$. The resulting
dreibein $\ce am = \rot_{ab} \ee bm$ will in general be complex, but the metric
$\gg^{mn}=\ce am \ce an =\ee am \ee an$ is still real and positive
semidefinite. If we now replace $\ee am$ by $\ce am$ in \eref{lag}, then we
obviously get an $\grp SO(3,\CC)$ invariant theory. All we have to do to keep
the metric real is to restrict the range of $\ce am$: only those values are
allowed that admit a rotation $\rot_{ab}\in\grp SO(3,\CC)$ such that $\ce am
\rot_{ab} \in \RR$, or alternatively we must have $\gg^{mn}$ positive
semidefinite. But this is a rather strange restriction, and the resulting range
is not a proper vector space or submanifold of $\CC^{3\times3}$, and so there
is no well-defined action principle any more.

To see this, let us examine what the allowed range for $\ce am$ looks like. We
know that there is an over-complete coordinate system on this space which is
given by specifying a real dreibein $\ee am\in\RR^{3\times3}$ and a complex
rotation $\rot_{ab}\in\grp SO(3,\CC)$, which makes $9+6=15$ real coordinates.
However, if we multiply $\rot_{ab}$ by an element of the $\grp SO(3,\RR)$
subgroup, we can compensate this be choosing a different real dreibein, so in
fact there are three ambiguous coordinates corresponding to the $\grp
SO(3,\RR)$ subgroup, and the space of the $\ce am$ has, at a generic point,
$12$ real dimensions only. Actually the complex dreibein is given by a real
dreibein and an element of the coset space $\grp SO(3,\CC)/\grp SO(3,\RR)$. A
suitable coordinate on the coset space is a real three-vector $v_a$, and a
possible standard representative is given by
\beq[boost3]
   \rot_{ab} = \alpha \, \delta_{ab} - (1+\alpha)^{-1} v_a v_b
               + \i \eps_{abc} \, v_c, \qquad  \alpha = \sqrt{1+v_av_a}.
\eeq
Let us call such an element of $\grp SO(3,\CC)$ a `boost' generated by $v_a$
(it is in fact the image of a boost in $\grp SO(1,3)_+$ under the group
isomorphism \eref{grp-iso}).
It is straightforward to verify that $\rot_{ab}\rot_{ac}=\delta_{bc}$ for any
$v_a\in\RR^3$, and that every group element can be written uniquely as a
product of some boost and a real $\grp SO(3)$ rotation. Hence, the coset is in
fact (topologically) an $\RR^3$. Together with the real dreibein we get $12$
real coordinates $(\ee am,v_a)$ for $\ce am$, and explicitly we can write
\beq[coor]
  \ce am = \rot_{ab} \ee bm =
          \alpha \, \ee am - (1+\alpha)^{-1} v_a v_b \ee bm
             + \i  \eps_{abc} \, v_c \, \ee bm.
\eeq
It is still not obvious what subset of $\CC^{3\times3}$ is covered by this map.
We can change coordinates from $(\ee am,v_a)$ to $(\re am,w_a)$ by setting
\beq[c-tr]
  \re am &=& \alpha \, \ee am - (1+\alpha)^{-1} v_a v_b \ee bm , \zl
  w_a &=& \alpha^{-1} v_a.
\eeq
The range of these new coordinates is $\re am$ arbitrary, but for $w_a$ we have
$w_aw_a<1$, i.e.\ it takes values inside the unit ball in $\RR^3$ only. We can
check that \eref{c-tr} is a proper change of coordinates, simply by giving the
inverse explicitly:
\beq
   \ee am &=& \alpha^{-1} \, \re am + \alpha (1+\alpha)^{-1} w_a w_b \re bm ,
    \zl
  v_a &=& \alpha w_a , \qquad \alpha = 1/\sqrt{1-w_aw_a}.
\eeq
Inserting the new coordinates into \eref{coor}, we find
\beq[coor2]
  \ce am =  \re am  + \i  \eps_{abc} \, w_c \, \re bm, \qquad
      w_a w_a <1.
\eeq
Hence, we got a rather simple decomposition of $\ce am$ into its real and
imaginary part. In contrast to this, \eref{coor} is rather a decomposition of
the complex dreibein into its modulus and phase (where the phase is the boost
which does not affect its modulus squared, i.e.\ the metric \eref{met-c}). We
can now see what the range of $\ce am$ is. It covers the whole real hyperplane
in $\CC^{3\times3}$, and at each point there is a three dimensional ball
attached which extends into the imaginary region. The orientation and radius of
the ball (or rather the axes of the ellipsoid) are linear functions of $\re
am$. The ball becomes degenerate if $\re am$ does. You can think of the
resulting range as some kind of angle, like, e.g., the spacelike region of a
Minkowski space including the origin. The latter is also given by attaching a
one dimensional `ball' to each point in the `$x_0=0$' plane of Minkowski space,
the (Euclidean) radius of that ball, i.e. its length, being the distance of its
center from the origin.

The resulting space is neither an open subset of $\CC^{3\times3}$ nor an open
subset of any submanifold, so we cannot choose $\ce am$ to be our basic field
variable. However, what we can do is to take either $(\re am,w_a)$ or $(\ee
am,v_a)$ as the basic set of fields. Then $\ce am$ is given as a composite
field by \eref{coor2} or \eref{coor}, respectively, and inserting this into
\eref{lag} leads to an $\grp SO(3,\CC)$ invariant Lagrangian
\beq[c-lag]
   \lag = \i \ce am \, \del_t A_{ma}
         +  \i A_{ta} \,   D_m \ce am
         -  \i N^m \, \ce an F_{mna}
         + \ft12 N \, \eps_{abc} \,\ce am \ce bn \, F_{mnc}.
\eeq
Note that with $\re am$ and $w_a$ as the basic fields, this is still
polynomial, but it does not share with \eref{lag} the property that $\re am$
and $A_{ma}$ can be treated as canonically conjugate variables in a
straightforward Hamilton Jacobi formulation, due to the fact that now each
component of $A_{ma}$ carries more than one real degree of freedom. Note also
that \eref{coor2} is not one-to-one, as for singular values of $\re am=\Re \ce
am$ we cannot necessarily recover $w_a$, so we cannot simply change to $\ce am$
as the basic field.

What we have now is a complexified version of Ashtekar's theory, which still
describes real gravity. The densitized spatial metric is given by
\beq[met-c]
  \gg^{mn} = \ce am \ce an = \ee am \ee an =
             (1-w_bw_b)\,\re am \re an + w_a \re am \, w_b \re bn.
\eeq
It is real and positive semidefinite, which follows from $w_aw_a<1$ if we
choose the last representation in terms of $\re am$ and $w_a$. Let us stick to
them, together with $A_{ma}$ and the multipliers $N$, $N^m$ and $A_{ta}$, as
the basic fields, and give an explicit representation for an $\grp SO(3,\CC)$
gauge transformation. In principle, we can start form $\ce am \mapsto \rot_{ab}
\ce bm$ and use the decomposition of $\grp SO(3,\CC)$ elements into boosts and
real rotation to find the transformation of $\ee am$ and $v_a$, then using
\eref{c-tr} to find those of $\re am$ and $w_a$. However, it is more convenient
to obtain the infinitesimal transformations form \eref{coor2} directly. What we
must have for the action \eref{c-lag} to be invariant is
\beq[c-a-tr]
   \delta \ce am = \eps_{abc} \,(\rot_c + \i\irot_c) \, \ce bm , \qquad
   \delta A_{ma} =  D_m (\rot_a + \i \irot_a),
\eeq
where the infinitesimal generator $\rot_a+\i\irot_a\in\alg so(3,\CC)\simeq
\CC^3$ has been split into its real and imaginary part. It is now
straightforward to verify that the nonlinear transformations
\beq[r-w-tr]
  \delta \re am &=& (\eps_{abc} \rot_c + w_d \irot_d\, \delta_{ab} -
                   w_a \irot_b ) \, \re bm , \zl
  \delta w_a &= & \irot_a + \eps_{abc} \rot_c w_b - \irot_b w_b w_a,
\eeq
inserted into \eref{coor2}, lead to \eref{c-a-tr}. We also see that $\delta
(w_a w_a) = 2 ( 1- w_a w_a) w_b\irot_b$ vanishes at the boundary of the unit
ball so that we cannot get out of the range of $w_a$ by a gauge transformation.

By choosing $\irot_a\propto -w_a$ and making a suitable finite gauge
transformation, we can always achieve $w_a=0$, which gives Ashtekar's theory
back as a gauge fixed version of \eref{c-lag}. Hence, it is not so obvious what
we actually got, beside a strange kind of complexification, which seems to be
more complicated than the original theory. However, now $A_{ma}$ is really an
$\grp SO(3,\CC)$ gauge field, and the action is still polynomial in all the
variables. When discussing the ADM formulation of the covariant theory in the
next section, it will be possible to transform the real four dimensional
vierbein directly into the real fields $\re am$ and $w_a$ by very simple
relations, and without any gauge fixing. Hence, the four dimensional Lorentz
group $\grp SO(1,3)_+$ can be directly identified with the $\grp SO(3,\CC)$
appearing here.

As a result, the three steps leading from `classical' Einstein gravity (in
vierbein formulation) to Ashtekar's theory \eref{lag}, namely extension to
degenerate metrics, ADM spacetime decomposition, and Lorentz gauge fixing, are
now completely separated. Moreover, we can perform these steps in any order,
thereby building up a cube of theories, Einstein and gauge fixed Ashtekar
gravity in two opposite corners, and the three space dimensions representing
the three steps. Two of these steps are rather technical, whereas extension to
degenerate metrics changes physics. We almost filled up all the corners of this
cube. With the results of this section, we got a non-gauge fixed version of
Ashtekar's gravity, which allows degenerate metrics. What is still lacking is
the theory that allows degenerate metrics but is manifestly diffeomorphism
invariant and leads to \eref{c-tr} by ADM decomposition and then to Ashtekar's
theory by gauge fixing. As this is technically the analog of Einstein gravity,
we expect it to provide the best insight into the `real physics' of Ashtekar's
gravity, as seen from the spacetime point of view. Though ADM gravity is useful
when considering questions like quantization or numerical computations, it is
the covariant formulation in terms of the 4-metric that `explains' the nature
of gravity. Hence, to understand the difference between Einstein and Ashtekar
gravity we need the last corner of the cube and it is this theory I want to
present in the following section.

\section{The covariant theory}\label{cov}
Let us forget Ashtekar's theory for a moment and start right from the beginning
by defining an extension of (real, Lorentzian) general relativity. We choose
the basic fields to be the $\alg so(1,3)$ spin connection $\O_{\mu AB}$ (with
field strength $R_{\mu\nu AB}$ defined in a straightforward way) and the
vierbein $\EE A\mu$, which transforms as a density of weight one half under
diffeomorphisms. As the action, we can take
\beq[act]
  \lag' = \ft12 \EE A\mu  \EE B\nu R_{\mu\nu}\^{AB},
\eeq
which is formally the same as in~\cite{bengtsson:89b}, and which may also be
considered as the vierbein version of the action used in~\cite{dragon:92}. It
defines an extension of Einstein gravity for degenerate metrics, corresponding
to singular matrices $\EE A\mu$.

\subsection*{Self-dual representation}
By introducing a complex basis of $\alg so(1,3)$, we can expand the spin
connection in terms of its self-dual and antiself-dual part (see the appendix
and~\cite{matschull:94a,matschull:95a} for the definition of the $J$ symbols):
\beq[O]
  \O_{\mu AB} = A_{\mu a} J_{aAB} + A\cc_{\mu a} J\cc_{aAB} ,
     \quad a=1,2,3.
\eeq
The basis $J_{aAB}=-J_{aBA}$ is orthonormal in the sense of \eref{ortho} and
provides the natural map of $\alg so(1,3)$ onto $\alg so(3,\CC)$.
The $\alg so(3,\CC)$ field strength is
\beq
   F_{\mu \nu a} =  \del_\mu A_{\nu a} - \del_\nu A_{\mu a} +
           \eps_{abc} A_{\mu b} A_{\nu c} = J_a\^{AB}  R_{\mu \nu AB}.
\eeq
Expanding the field strength like \eref{O}, $\lag'$ becomes the real part of
\beq[act-sd]
   \lag =  \EE A\mu  \EE B\nu J_a\^{AB} F_{\mu\nu a}.
\eeq
This provides an extension of Einstein gravity for singular metrics as well,
which is slightly different from $\lag'$. It is $\lag$ that will be equivalent
to Ashtekar's gravity. The difference between the two extensions is not
important for our purpose here, but it is quite interesting. The equations of
motion for the spin connection are identical, because $\lag$ is holomorphic in
$A_{\mu a}$ and therefore becomes stationary if and only if its real part is
stationary. However, $\lag$ provides an additional field equation for the
vierbein. We find
\beq[eom]
         {{\delta\lag} \over {\delta\EE A\mu}} &=& 2 \,
                                 \EE B\nu J_a\^{AB} F_{\mu\nu a} = 0
            \zl&&\Leftrightarrow\qquad
        \EE B\nu  R_{\mu\nu}\^{AB} = 0 , \qquad
        \EE B\nu \eps^{ABCD}  R_{\mu\nu CD} = 0 .
\eeq
The last two equations are obtained by taking the real and imaginary part of
the first equation. The first one is the Einstein equation, which is also the
equation of motion for $\EE A\mu$ in $\lag'$. The second equation is not
implied by $\lag'$. For invertible metrics, it is the first Bianchi identity
for the Riemann tensor $R_{\mu[\nu\rho\sigma]}=0$, written in vierbein
formulation. For singular vierbein fields this is in general not a consequence
of the remaining equations of motion. So the self-dual formulation $\lag$
additionally imposes the Bianchi identity for the curvature tensor as an
equation of motion, whereas $\lag'$ does not.

To see the relation between the Lorentz group $\grp SO(1,3)_+$ and its
self-dual representation $\grp SO(3,\CC)$, let us consider gauge symmetries of
the self-dual action. A finite Lorentz transformation is given by a real
$4\times4$ matrix $\Rot_A\^B\in\grp SO(1,3)_+$ (obeying
$\Rot_A\^C\Rot_B\^D\eta_{CD}=\eta_{AB}$, $\det(\Rot)=1$, and $\Rot_0\^0>0$), or
equivalently by a complex $3\times3$ matrix $\rot_{ab}\in\grp SO(3,\CC)$
(obeying $\rot_{ab}\rot_{ac}=\delta_{bc}$ and $\det(\rot)=1$). Thereby, the
fields have to transform such that
\beq
   \EE A\mu \mapsto \Rot_A\^B \EE B\mu ,\qquad
   F_{\mu\nu a} \mapsto \rot_{ab} F_{\mu\nu b}.
\eeq
To make this a symmetry of the action, we must have
\beq[grp-iso]
   \rot_{ab} = J_a\^{AB} J_{bCD} \Rot_A\^C \Rot_B\^D =
               - \Rot_{00} \Rot_{ab} + \Rot_{a0} \Rot_{0b}
               + \i \eps_{bcd} \Rot_{ac} \Rot_{0d},
\eeq
which follows by direct computation from the properties of the $J$ symbols
given in the appendix. This is the natural map $\grp SO(1,3)_+\to\grp
SO(3,\CC)$, which can be checked to be a one-to-one group homomorphism (there
is no sign ambiguity due to the quadratic form, because $\Rot_{00}<0$). It
obviously maps the $\grp SO(3,\RR)$ subgroups identically onto each other.
Hence, we already have the $\grp SO(3,\CC)$ representation of the Lorentz group
on the four dimensional level, $A_{\mu a}$ being the corresponding gauge field.
To get Ashtekar's gravity after a spacetime decomposition, all we have to do is
to identify the vierbein components with the fields appearing in \eref{c-lag},
or after a gauge fixing with those appearing in \eref{lag}.

\subsection*{ADM formulation}
In Einstein gravity, the ADM formulation is possible only for special spacetime
manifolds (those that admit a slicing $\M=\N\times\RR$), and a given slicing
also restricts the space of field configurations, because the slices have to be
spacelike everywhere. Expressed in terms of the scalar function $T(\X)$
defining the slicing, this means that the covector orthogonal to the slices
($\del_\mu T$) has to be timelike ($\del_\mu T\,\del_\nu T\,G^{\mu\nu}<0$),
which imposes a restriction on the (invertible) metric $G_{\mu\nu}$.
As a consequence, the ADM formulation is no longer manifestly covariant under
four dimensional diffeomorphisms, and it also excludes some field
configurations, e.g.\ those with closed timelike curves. However, for any given
field configuration such that the slices are spacelike, we can always change
the slicing `slightly' in any direction, such that the new slices are still
spacelike. So the ADM formulation does allow `small' diffeomorphisms, which can
also be checked by computing the transformations of the fields under
infinitesimal diffeomorphisms. When the diffeomorphisms become `too big', then
these infinitesimal transformations can no longer be integrated. In Einstein
gravity we can easily say when this is going to happen. Namely, when the slices
become lightlike at some point.
The aim of this section is now to make an analogous and straightforward
construction for the covariant theory defined by \eref{act-sd}, leading to
Ashtekar's gravity as its ADM formulation.

An important notion in ADM formulation of Einstein gravity is that of a
spacelike hypersurface. It has a straightforward generalization for degenerate
metrics. Let us call a surface spacelike if its normal covector $\W_\mu$ is
timelike in the sense that $\W_\mu\EE A\mu$ is a timelike vector in Minkowski
space, or $\W_\mu\,\W_\nu\,\GG^{\mu\nu}<0$. An important feature of this
definition is that it allows `small' deformations, as described above for
invertible metrics. Consider a given metric $\GG^{\mu\nu}$ at some fixed point
in spacetime. Then, the set of all normal vectors $\W_\mu$ satisfying
$\W_\mu\,\W_\nu\,\GG^{\mu\nu}<0$ is obviously a (possibly empty but) open
subset of $\RR^4$. Therefore, a spacelike surface can be deformed slightly in
any direction, thereby remaining spacelike. Anticipating the result that
Ashtekar's theory is the ADM formulation of our covariant theory, this is in
agreement with the fact that there are well defined infinitesimal
transformation \eref{infi} for Ashtekar's theory. The fact that these might not
be integrable for `too big' diffeomorphisms means that the slices become
`non-spacelike' (there is no straightforward generalization of `lightlike' or
`timelike' surfaces).

For degenerate vierbeins $\EE A\mu$ strange things can happen: it might be that
at a given point there is no spacelike hypersurface at all. This happens when
the image of $\EE A\mu$, viewed as a linear map from the cotangent space of
spacetime into the four dimensional Minkowski space, does not contain any
timelike vector, so for example in the trivial case $\EE A\mu=0$. Hence, there
are different kinds of degeneracy, and we will see that due to this there is a
crucial difference between the relation of Einstein gravity to its ADM
formulation and the relation between our covariant theory and Ashtekar's
gravity. In this sense \eref{act-sd} is not yet what we should call the
covariant version of Ashtekar's gravity. I will come back to this problem
below.  Here, this `worst' kind of degeneracy will be ruled out simply by
imposing the usual ADM restriction: for a given slicing $T(\X)$ of $\M$, we
restrict the range of the vierbein such that for all $\X\in\M$
\beq[timelike]
  \EE A\mu(\X) \,\del_\mu T(\X)  \txt{is negative timelike.}
\eeq
A vector $\W^A$ is called negative timelike if $\W^A\W_A<0$ and $\W^0<0$. This
means that in addition to requiring the slicing to be spacelike we also fix the
sign of $\EE A\mu$, which in some sense means that $T$ increases when going
towards the physical time direction. It does not restrict the set of solutions,
because with $\EE A\mu$, $-\EE A\mu$ solves the field equations as well, and we
just have to replace $T\mapsto-T$ to find the solutions excluded by the word
`negative' in \eref{timelike}. In the next section I will give a more precise
definition of how the physical arrow of time is assumed to be included in the
the vierbein field, and you can check that `negative' in \eref{timelike}
implies that $T$ increases when going towards the future. Of course,
\eref{timelike} also excludes the case that there are points where no spacelike
hypersurface exists.

Introducing the coordinate $t=T(\X)$ and three more (possibly local)
coordinates $x^m$, the action \eref{act-sd} splits into
\beq[act-adm]
   \lag =  2 \, \EE At  \EE Bm J_a\^{AB} F_{tma} +
             \EE Am  \EE Bn J_a\^{AB} F_{mna}.
\eeq
Now there are several possible ways to proceed. Because of \eref{timelike}, we
now have $\EE At$ negative timelike, so we can find a boost $\Rot_A\^B\in\grp
SO(3,\CC)$ such that the rotated vierbein is of the form \eref{E-def}, with
positive $\EE 0t$ (note the position of the $0$ index). Hence, we can impose a
gauge fixing and it is straightforward to verify that \eref{act-adm} becomes
the same as \eref{lag}.

If we want to avoid any kind of gauge fixing, we must find a transformation
from \eref{act-adm} to \eref{c-lag}, i.e. we have to define the fields $N$,
$N^m$, $\re am$ and $w_a$ appearing therein as a function of $\EE A\mu$ such
that
\beq
      2 \EE At  \EE Bm J_a\^{AB} &=&  \i\ce am, \zl
        \EE Am  \EE Bn J_a\^{AB} &=&  \ft12 N \eps_{abc} \,\ce bm \ce cn
                                     - \i N^{[m} \ce a{n]}.
\eeq
By writing out real and imaginary parts of these equations and using the
explicit representation \eref{J-def} for the $J$ symbols one finds the very
simple relations
\beq[fd-map]
  N^{-1} = - \EE at \EE at +  \EE 0t \EE 0t , &\qquad&
  \re am = \EE at \EE 0m - \EE 0t \EE am, \zl
  N^{-1} N^m =  \EE at \EE am - \EE 0t \EE 0m , &\qquad&
  w_a = \EE at / \EE 0t .
\eeq
Of course, $N^{-1}=-\GG^{tt}$ and $N^{-1}N^m=\GG^{tm}$, I just wrote out the
components of the vierbein to show the similarity of these formulae. Note that
\eref{timelike} ensures that $w_aw_a<1$ and $N>0$. The relations are
invertible, as for a given set $N,N^m,\re am,w_a$ we can recover $\EE At$ from
$N$ and $w_a$. The remaining equations are linear in $\EE Am$, and it is easy
to check that they have a unique solution. Hence, the somewhat unmotivated
introduction of the `canonical' variables $\re am$ and $w_a$ in \eref{c-tr} not
only leads to the decomposition \eref{coor2} of the complex dreibein. It also
gives this surprisingly simply transformation from the vierbein to the three
dimensional variables. If $\EE at=0$, i.e.\ if we impose a gauge fixing, we
have $w_a=0$, $\re am=\ee am$, and the three dimensional fields are given by
\eref{E-def}.

If you prefer the variables $\ee am$ and $v_a$ to parameterize the complex
dreibein $\ce am$ in \eref{c-lag}, you can proceed as follows. Given the
vierbein $\EE A\mu$, subject to \eref{timelike}, find the (unique) boost
$\Rot_A\^B$ that brings the vierbein into the upper triangular form. Then you
get $\ee am$ by \eref{E-def}, and $v_a$ is obtained by mapping the boost into
$\grp SO(3,\CC)$, which is then of the form \eref{boost3}.

As a result, we found a manifest covariant theory of gravity \eref{act-sd},
whose ADM formulation is Ashtekar's gravity. However, there remains one crucial
point to be considered which makes this ADM formulation different from that of
Einstein gravity. It is the fact that in our theory the metric might be
degenerate in such a way that at some point no spacelike hypersurface exists.
Such a field configuration has no representation in the ADM formulation, and
therefore no representation in Ashtekar's theory. This is a well know problem
in Einstein gravity too, as there are many field configuration which do not
have representations in ADM gravity, e.g.\ those with closed timelike curves
etc. But there is a difference: in our case, the problem is {\em local},
whereas the obstacles arising in the ADM formulation of Einstein gravity are
always of a global type. In some sense, Einstein gravity is {\em locally\/}
equivalent to its ADM formulation, whereas our covariant theory is not yet
locally equivalent to Ashtekar's theory.

\subsection*{Local slicing}
To make this more precise, let us define what is meant by `local equivalence'
of Einstein gravity and its ADM formulation. Let $G_{\mu\nu}$ be a field
configuration of Einstein gravity, i.e.\ an invertible, differentiable,
Lorentzian metric on some spacetime $\M$, and let $\X_0\in\M$. Choose any
timelike covector $\V_\mu(\X_0)$ at $\X_0$, and define a scalar field $T(\X)$
such that $\del_\mu T(\X_0)=\V_\mu(\X_0)$. As the metric is continuous, there
will be a neighbourhood of $\X_0$ where $\del_\mu T$ is timelike, and thus also
$\del_\mu T \neq 0$. Inside this neighbourhood $T$ defines a `local slicing'.
One can introduce the ADM variables, and express the contribution of this
volume element to the action in terms of the ADM action. The field equations
derived from this action are equivalent to those from the Einstein Hilbert
action, and hence, at least locally, one can always go over to the ADM
formulation of gravity. Only when requiring that a {\em global} spacetime
slicing exists, we get a restriction on the field configurations.

This is not true for our theory. If there is no spacelike hypersurface at
$\X_0$, then there is no local slicing $T(\X)$, and the transition to
Ashtekar's variables and finally the action \eref{lag} is not possible. We
somehow have to restrict the range of $\EE A\mu$, but in a fully covariant way,
i.e.\ not referring to any coordinate system, to get a theory which is {\em
really\/} the covariant version of Ashtekar's theory. It is not so difficult to
guess how this restriction has to look. For each $\X\in\M$ we must have:
\beq[condx]
   \exists \V_\mu(\X) \txt{such that}
     \V_\mu(\X) \EE A\mu(\X) \txt{is negative timelike.}
\eeq
This $\V_\mu$ is the normal covector of some hypersurface, which is spacelike
by definition. So we can equivalently require that
\beq[cond]
  \hbox{there is a spacelike hypersurface at each point in spacetime.}
\eeq
In contrast to \eref{timelike}, this does no longer refer to any global object
like the scalar field $T$ or any coordinate. With \eref{cond}, our theory is
still covariant and an extension of Einstein gravity, as it is certainly
fulfilled by invertible vierbeins. Another question is whether the action
principle is still well defined, which requires that the set of allowed values
for $\EE A\mu$ does not have boundaries. It is in fact open (as a subset of
$\RR^{4\times4}$). For every $\EE A\mu$ with $\GG^{\mu\nu}\V_\mu\V_\nu<0$ for
some $\V_\mu$, all vierbeins solving this inequality (for the same $\V_\mu$)
have the required property, and they obviously form an open neighbourhood of
$\EE A\mu$.

With the restriction \eref{cond}, we can now show that our covariant theory is
locally equivalent to Ashtekar's theory, by the same arguments as before for
Einstein gravity. Starting at some point $\X_0$, there is a negative timelike
covector $\V_\mu(\X_0)$. Define $T(\X)$ such that $\del_\mu
T(\X_0)=\V_\mu(\X_0)$, then there is a neighbourhood where \eref{timelike}
holds, the transition to the variables \eref{fd-map} can be performed, and the
contribution to the action from this volume element reads \eref{c-lag}.

To summarize, the action \eref{act-sd} together with the restriction
\eref{cond} provides a covariant version of Ashtekar's gravity, allowing
exactly the right kinds of singular metrics. There are two crucial points where
the theory is different from that in~\cite{bengtsson:89b}: first of all, here
we are dealing with real gravity, there is no need to complexify the metric,
and the gauge group is the four dimensional Lorentz group $\grp
SO(1,3)_+\simeq\grp SO(3,\CC)$. No gauge fixing is necessary to transform to
the three dimensional  variables in the complexified version of Ashtekar's
gravity, except for the `global' ADM formulation where the same restriction
(`spacelike slices') as in Einstein gravity is required, and which can (partly)
be understood as a gauge fixing. The second difference is that it was possible
to render the condition $N\neq0$ covariant, which in~\cite{bengtsson:89b}
somehow remains as a non-covariant relic. What remains to be done now is to
consider the covariant theory from a physical point of view.

\section{The causal structure}\label{caus}
In this section I want to analyse the geometrical properties of the covariant
theory, at a kinematical level, i.e.\ only considering the field configurations
themselves but not the dynamics. This will give us a deeper understanding of
what the singular metrics in Ashtekar's theory actually are, in particular we
will see how spacetime looks like `at the origin' $\ee am=0$. We will find
that, even in this highly degenerate case, there is (at least locally) a well
defined causal structure. This is in contrast to a comment made by
Bengtsson~\cite{bengtsson:91}, who assumes that degenerate metrics describe
spacetimes without causal structure. So it is quite remarkable that Ashtekar's
singular metrics are {\em exactly\/} those which {\em do not\/} destroy the
causal structure.

On the other hand, this is not really surprising because otherwise the Hamilton
Jacobi formalism would not work. Nevertheless it is interesting to be
considered from the four dimensional point of view. In contrast to Einstein
gravity, where in some sense the `origin' of the fields is flat spacetime, here
this origin turns out to be a space consisting of a continuum of completely
disconnected points, but with a well defined `time' everywhere. The same kind
of origin is considered by Dragon~\cite{dragon:92}, but there the time
direction becomes degenerate, too: not only the space points but also the
spacetime events are completely disconnected. Hence, our theory is a little bit
more `physical' than Dragon's, and an interesting fact is that a distinction
between `time' and `space' is made without destroying the manifest invariance
under diffeomorphisms, which is due to the notion `spacelike' appearing in
\eref{cond}.

\subsection*{Future, past, and spacelike hypersurfaces}
Given an arbitrary vierbein, not necessarily subject to \eref{cond}, you may
consider $\EE A\mu(\X):\MM^4\to T_\X\M,\,\W^A\mapsto\W^\mu(\X)=\W^A\EE
A\mu(\X)$ as a map from four dimensional Minkowski space into the tangent
bundle of the spacetime $\M$ (actually from an $\MM^4$ bundle over $\M$, which
only has to be time orientable for our purpose, but let us for simplicity
assume that it is trivial). Let us define the `future' of a point $\X\in\M$ as
the set of all tangent vectors at $\X$ which are images of positive timelike or
lightlike vectors in $\MM^4$, i.e.
\beq[future]
  \fut(\X) = \big\{ \W^\mu(\X)=\W^A\EE A\mu(\X) \,\big|\,
                      \W^A\W_A \le 0 , \W^0 > 0 \big\} .
\eeq
The past is defined similarly as $-\fut(\X)$, and a causal curve is
straightforwardly defined as a curve whose tangent vector is contained in the
future at every point the curve passes through.
The time orientation is assumed to be something real physical, so that the
gauge group is really $\grp SO(1,3)_+$ and changing the sign of $\EE A\mu$ is
not considered as a gauge transformation, because it changes the time
direction. If you like you can choose $\grp O(1,3)_+ \simeq \grp O(3,\CC)$ to
be the gauge group, allowing spacelike parity transformations. Hence, in
addition to the metric, the vierbein is assumed to carry information about the
physical arrow of time as well.

The shape of the future can be very different for different values of the
vierbein. If $\EE A\mu(\X)$ is invertible, it is obviously the usual future
lightcone together with its interior at $\X$ (but $\W^\mu=0$ excluded). Let us
call this a hypercone, as it has one dimension more than a usual cone. If the
rank of $\EE A\mu(\X)$ is less than 4, the future becomes degenerate, and the
question to be considered here is whether such a degenerate future still has
the physical features it should have. There are three possible ways in which a
hypercone can become degenerate under a rank 3 linear map. Depending on which
direction it is  projected into, it either becomes a (three-dimensional) cone,
with the peak excluded, a hyperplane, or a half-hyperplane including the
boundary. The latter happens if it is projected along a lightlike direction.
{}From a physical point of view, the last two cases are worse than the first.
In the first case, we have one space direction in which light does not
propagate (i.e.\ its velocity vanishes), but it propagates normally into two
other directions, and the future somehow still looks like a lightcone, except
that it is `a bit' flattened. However, if the future becomes a hyperplane,
there is a direction in which light propagates infinitely fast, and, moreover,
the future will intersect with, or even becomes equal to the past. Hence, the
causal structure of spacetime is lost. A similar situation occurs for smaller
rank of $\EE A\mu$: for rank 2 we get an angle, the peak excluded, a half-plane
or a plane, and for rank 1 the future becomes a half line with or without end
point, or a full line. For $\EE A\mu=0$, finally, the future is $\{0\}$.

Now, which of these situations are physically reasonable and which are not?
What we want is that the future somehow `points into a direction' and the past
points into the opposite direction. In particular, this means that the future
is non-trivial and does not intersect with the past. So let us impose the
following condition on the vierbein
\beq[cond2]
   -\fut(\X) \cap \fut(\X) = \emptyset
     \quad \Leftrightarrow \quad 0 \not\in\fut(\X).
\eeq
The equivalence of the two conditions is easy to see. $\Rightarrow:$ if
$0\in\fut$, then clearly $0\in-\fut$. $\Leftarrow:$ if there is some
$\W^\mu\in-\fut\cap\fut$, then we have $\W^\mu\in\fut$ and $-\W^\mu\in\fut$,
but with two vectors their sum is an element of $\fut$ as well, which follows
immediately from \eref{future} and the fact that the sum of two positive
timelike or lightlike vectors in Minkowski space is such a vector again. So
$0\in\fut$.

Condition \eref{cond2} also implies that the future is non-trivial: by
definition, it cannot be empty, so there is at least one non-zero tangent
vector in $\fut(\X)$. If \eref{cond2} holds, the future is either a hypercone,
a cone, an angle, or a half-line, depending on the rank of the vierbein. In all
these cases there is a at least one causal curve through $\X$. It is only the
number of the degrees of freedom of this curve which is affected by the rank of
$\EE A\mu$. Moreover, the fact that future and past do not overlap means that
there is space in between for, e.g., a hyperplane that does not intersect with
either of them. For invertible metrics such a surface is spacelike. It is
reasonable to call a surface spacelike if it intersects neither with the future
nor with the past, as it then contains causally disconnected points only. Let
us check whether this definition of a spacelike surface is equivalent to that
given in the last section, and whether the condition \eref{cond2} coincides
with \eref{timelike}. If this is the case, we have a new definition for the
range of the vierbein in our theory, which is somehow more physical.
To proof the equivalence, we can use a simple {\em Lemma:} the following two
statements are equivalent:
\begin{itemize}
\item[(a)] $\V_A\in \MM^4$ is negative timelike, and
\item[(b)] $\V_A\W^A>0$ for all positive timelike or lightlike $\W^A\in\MM^4$.
\end{itemize}
To proof this, you just have to write out the scalar product in components: for
$(a)\Rightarrow(b)$ use that $\V^0<-|\V_a|$, $\W^0\ge|\W_a|$, and the Schwarz
inequality; for $(b)\Rightarrow(a)$ choose $\W^A=(|\V_a|,-\V_a)$, which is
positive lightlike and gives $\V_A\W^A>0\Rightarrow\V^0<-|\V_a|$.

If we now consider the two definitions for spacelike surfaces, it is easy to
proof their equivalence: a surface is spacelike if
\begin{itemize}
\item[(a)] the surface does not intersect with $\fut$, or
\item[(b)] the normal covector $\V_\mu$ can be chosen such that $\EE A\mu
\V_\mu$ is negative timelike.
\end{itemize}
Let $\V_\mu$ be the normal covector of the surface, then (a) means that
$\V_\mu\W^\mu\neq0$ for all $\W^\mu\in\fut$. As $\fut$ is connected, we can
choose $\V_\mu$ such that the sign is positive, i.e.\ $\V_\mu\W^\mu=\V_\mu\EE
A\mu\W^A>0$ for all positive timelike or lightlike $\W^A$ (this is the
definition of $\fut$). Using the Lemma, this is equivalent to (b), stating that
$\V_\mu\EE A\mu$ is negative timelike.

What remains to be shown is that \eref{cond2} is equivalent to requiring that
there is a spacelike hypersurface at any point in spacetime. Assume that there
is such a surface, then certainly $0\not\in\fut$, because it lies inside the
surface. On the other hand, if no spacelike hypersurface exists, then for every
covector $\V_\mu$ we have $\E A\mu\V_\mu$ spacelike or lightlike (otherwise
this would be a normal covector of some spacelike hypersurface). So all these
vectors lie on a non-timelike hyperplane in Minkowski space, having a positive
timelike or lightlike normal vector $\W^A$. Hence, we have $\W^A\EE
A\mu\V_\mu=0$ for all $\V_\mu$, i.e.\ $\W^A\EE A\mu=0$. But by definition this
is an element of the future, so $0\in\fut$.

We now have a `physical' condition \eref{cond2}, which defines the kinematics,
i.e\ the set of allowed vierbein fields, of the covariant version of Ashtekar's
gravity. The result is that only those degenerate metrics are allowed which
provide a local causal structure, in the sense that there is a subset of the
tangent space at each point which denotes the future. This is separate from the
past, pointing towards the opposite direction, and in between there is place to
put hypersurfaces which do not intersect with the future (nor the past), and
those are called spacelike. Another equivalent restriction on the vierbein is,
as already guessed in the first section, that the metric $\GG^{\mu\nu}$ must
have signature $(-,+,+,+)$, $(-,+,+,0)$, $(-,+,0,0)$ or $(-,0,0,0)$. I will not
give an explicit proof for this, but you can infer from \eref{condx} that there
is at least one `$-$' sign, and the signature of $\eta_{AB}$ implies that there
cannot be more `$-$' signs in the signature of $\GG^{\mu\nu}$. In principle,
what all these equivalent conditions say is that space may become degenerate
but not time. To illustrate the situation of a degenerate future, let us
consider some simple examples.

\subsection*{Scalar fields and black holes}
As we are only interested in the kinematics of degenerate metrics, let us
consider a scalar field in a given background, i.e.\ with fixed values for the
vierbein $\EE A\mu$ or the metric $\GG^{\mu\nu}$. Its action reads
\beq
   \Lmat = - \ft12 \int \d^4\X \, \GG^{\mu\nu}
           \, \del_\mu \varphi \, \del_\nu \varphi .
\eeq
Note that the simple structure of this action was one of the motivations
in~\cite{dragon:92} to use $\GG^{\mu\nu}$ as the basic gravitational field
(beside some regularity requirements which are not so different from Ashtekar's
`polynomiality'). As mentioned there, degeneracy of the metric means that there
are directions in spacetime in which $\varphi$ can fluctuate without giving
contributions to the energy, or equivalently without influencing the equations
of motion. For a completely degenerate metric $\GG^{\mu\nu}=0$, the field
$\varphi$ drops out from the action and evolves arbitrarily. In our case,
however, there is always at least one direction where the metric is not
degenerate, and this direction is timelike. Given a causal curve through some
point in spacetime, then there is always a positive contribution to the
`kinetic' energy of $\varphi$ in the action above, which is quadratic in $\dot
\varphi$, where the dot denotes the derivative with respect to the curve's
parameter. As a consequence, the equation of motion can always be solved for
$\ddot\varphi$, and we always have a unique time evolution if we follow a
causal curve through spacetime.

In the `most degenerate' case, the future consists of a half line only at every
point in spacetime. The metric takes the form $\GG^{\mu\nu}=-\W^\mu\W^\nu$,
where $\W^\mu$ is some vector (of weight one half) in $\fut$, which itself
consists of the positive multiples of $\W^\mu$ only. This corresponds to the
`origin' $\ee am=0$ of Ashtekar's canonical variables. Let us assume that
spacetime consists, at least locally, of a bundle of non-intersecting causal
curves, such that exactly one of them passes through each point (you can find
some strange vector fields $\W^\mu$ without this feature, but those do not
admit a local slicing such that $\ee am=0$). Then, we can find a coordinate (or
slicing) $t$ such that $\W^\mu=(1,0,0,0)$, and the action for the scalar field
becomes
\beq[Lmat]
  \Lmat = - \ft12 \int \d^ 3x \int\d t \, (\dot\varphi)^2.
\eeq
Obviously, this describes a continuum of completely decoupled fields
$\varphi(x)$ for each `space' point $x$ (if the slicing is global). Physically,
space consists of a set of completely disconnected points, each having its own
value of $\varphi$ evolving in time, thereby ignoring what is going on in the
neighbourhood. Another way to see this, which will be useful below, is to
consider a field that has a `step', and to check whether this step contributes
to the energy or not. So we set $\varphi = \varphi_0 + \alpha\t$, where
$\varphi_0$ is continuous, $\alpha$ some number, and $\t$ the characteristic
function of some region of space, which is one inside and zero outside that
region. If we plug this into the action, we find that, whatever the region is,
the action does not depend on $\alpha$ and therefore there is no contribution
to the energy coming from the step along the boundary of the specified region.
This is what is meant by `a fluctuation does not contribute to the energy'
above, and tells us that any given region and its complement are physically not
connected, i.e.\ no information (about $\varphi$) can be passed from one point
to another.

Defining this situation as the natural origin of the gravitational fields,
which one usually does in most of the applications of Ashtekar's variables,
gravity is no longer the `deviation' from flat spacetime. Instead, switching on
the fields somehow switches on the communication between adjacent points in
space, so that they know of each other and information can be passed from one
to another. Flat spacetime is just a very special configuration. In a
perturbative formulation this leads to a new understanding of what a graviton
is. Instead of being a perturbation of flat space, you should consider it as a
deviation form the completely `discretized' space described above. Flat space
would no longer be the vacuum but some kind of condensate in perturbative
quantum gravity, and because we are expanding around a very different classical
field configuration, this might have some influence on renormalizability and
related problems in perturbative quantum gravity, which so far have not been
investigated in the context of Ashtekar's variables.

Let us also consider a special field configuration where the metric is
`slightly' degenerate only. With global coordinates $t,x,y,z$, we take
\beq[mm]
  \GG^{\mu\nu} = \diag(-1, 1 , 1 , f^2(z) ),
\eeq
where $f$ is some function with $f(0)=0$, $f'(0)=1$ and $f(z) \to 1$ for
$z\to\pm\infty$, so that spacetime is asymptotically flat for large $z$ and the
metric is singular on the $z=0$ plane. A possible choice is
$f(z)=z/\sqrt{1+z^2}$, but we won't need $f$ explicitly. The action for the
scalar field becomes
\beq
  \Lmat = -\ft12 \int \d^4\X \big( \nabla \varphi \cdot \nabla \varphi
                            + f^2(z) \, \del_z \varphi\,  \del_z \varphi \big),
\eeq
where $\nabla$ denotes the derivative operator $(\del_t,\del_x,\del_y)$ with
respect to the `flat' coordinates. To see what happens at the `wall', we again
consider a field which has a finite step, i.e.\ which is of the form
$\varphi=\varphi_0+\alpha\t(z)$, where $\varphi_0$ is continuous, and $\t(z)$
is the step function taking the values $0$, $1$ for $z<0$, $z\ge0$,
respectively. As $\del_z\t(z)=\delta(z)$, this unfortunately produces a
$\delta^2(z)$ term in the action, so we have to regularize somehow. We do this
by defining a regularized step function
\beq
  \t(z)=\cases{0 & for $z\le-\epsilon/2 $,\cr
               z/\epsilon+\ft12 & for $-\epsilon/2 \le z \le \epsilon/2$, \cr
               1 & for $z\ge\epsilon/2$,}
\eeq
and taking the limit $\epsilon\to0$. The contribution of the step to the
kinetic energy then becomes
\beq[stac]
    - \frac\alpha\epsilon \int_{-\epsilon/2}^{\epsilon/2} \!\!\!\d z
             \, f^2(z) \, \del_z \varphi_0
    - \frac{\alpha^2}{2\epsilon^2} \int_{-\epsilon/2}^{\epsilon/2} \!\!\!\d z
             \, f^2(z).
\eeq
Using $|f(z)|\le |z|$ and the fact that $\varphi_0$ is continuous it is easy to
show that this vanishes in the limit $\epsilon\to0$. Hence, we see that the
fields on both sides of the wall decouple completely and there is no
correlation.

Another way to see this is to look at light rays approaching the wall. For
$z>0$, the metric is invertible and we have $G_{\mu\nu}=\diag(-f,f,f,f^{-1})$.
Hence, for a light ray in the $t,z$ plane we have $\d t/\d z=\pm f^{-1}(z)$, so
that $t$ diverges logarithmically for $z\to0$, and the light never reaches the
wall, showing again that there is no communication between the two parts of
spacetime. Of course, all these arguments depend crucially on the behaviour of
$f$ near $z=0$. Here we assumed that $f(z)\approx z$. If $f$ behaves
differently, we get very different kind of `walls'. Without giving proofs
(which are absolutely straightforward), let me just list a few `physical'
properties such a wall can have:
\begin{itemize}
\item[(1)] if $z f^{-2}(z)$ is bounded for $z\to0$, then the scalar fields on
both sides do not decouple (in this case \eref{stac} does not to vanish);
\item[(2)] if $z f^{-1}(z)\to0$ for $z\to0$, light rays from outside can touch
the wall (because $f^{-1}(z)$ is integrable at $z=0$);
\item[(3)] if $z f^{-1/2}(z)\to0$ for $z\to0$, the spacelike distance between
the wall and some point outside is finite.
\end{itemize}
We see that (1) implies (2) and (2) implies (3), which is reasonable from the
physical point of view, too. But note that they are not equivalent. So already
with this rather simple degenerate metric \eref{mm} we can describe very
different singular structures in spacetime, in this case some kind of `domain
walls'. Moreover, the singular metrics somehow enable us to glue together parts
of spacetime which do not communicate with each other, i.e.\ they naturally
describe horizons, which are however stronger than usual horizons which only
exist with respect to specific observers (remain `outside' for all times), and
where information can be passed through in one direction. Remember, however,
that we only considered the kinematics here, i.e.\ it is not clear whether all
these structures may appear dynamically, i.e.\ as solutions of the full set of
field equations.

A typical situation where such a wall in fact occurs dynamically is the
Schwarzschild geometry, when simple polar coordinates $t,r,\theta,\varphi$ are
used. There we have a coordinate singularity at the horizon $r=r_0$, which is
similar to the $z=0$ wall above. It has been shown
in~\cite{bengtsson:93,bengtsson:91} that Ashtekar's variables can be used to
avoid this singularity, i.e.\ one can choose coordinates such that all the
fields are finite at the horizon. Those are quite different from Kruscal or
related coordinates, which eliminate the singularity by a non-regular
coordinate transformation. Instead, the transformation here is invertible, and
there is still a degenerate metric at the horizon. Moreover, one finds very
peculiar solutions of the field equations, called `empty black holes', i.e.\
the metric inside the black hole is not of the Schwarzschild type but simply
flat spacetime, with only a coordinate singularity at the origin.

With our results from above these solutions are no longer mysterious, as the
`wall' keeps the part of spacetime behind the horizon separate from that
outside the black hole, so there is no relation between the `physics' inside
and outside the $r=r_0$ sphere. In fact, the region $r<r_0$ is not really the
interior of the black hole. The latter can be found somewhere else, namely by
transforming to, say, Kruscal coordinates and extending spacetime behind the
true horizon. Bengtsson's `empty black hole' is therefore rather a combination
of two different, and separated, spacetimes, glued together along a wall
similar to the $z=0$ wall above, but this time the wall has the shape of a
sphere.

\section*{Acknowledgements}

I would like to thank Michael Reisenberger for drawing my attention to a
mistake made in~\cite{matschull:94a}, which led to the wrong conclusion that
diffeomorphisms are not well defined for singular values of Ashtekar's
variables. This paper was intended to correct the mistake and to clarify the
problem of finite diffeomorphisms in Ashtekar's gravity, at the classical
level.
\def\theequation{A.\arabic{equation}}
\section*{Appendix}
This is a collection of formulae for the $J$ symbols, for more detailed
information see~\cite{matschull:94a,matschull:95a}.
\noindent Explicit representation ${\ssty A},{\ssty B},\dots = 0,1,2,3$,
$a,b,\dots=1,2,3$):
\beq[J-def]
J_{aAB} = \ft\i2 \eta_{aA} \delta_B\^0 - \ft\i2 \eta_{aB} \delta_A\^0
           - \ft12 \eps^0\_{aAB},
\eeq
Orthonormality and completeness:
\beq[ortho]
&&J_a\^{AB} J_{bAB} = \eta_{ab}, \qquad
J\cc_a\^{AB} J\cc_{bAB} = \eta_{ab}, \qquad
J_a\^{AB} J\cc_{bAB} = 0 , \zl
&&J_a\^{AB} J_{aCD} + J\cc_a\^{AB} J\cc_{aCD} =
  \delta^A\_{[C} \delta^B\_{D]}.
\eeq
Commuting representations of $\alg so(3)$:
\beq[so3-rep]
J_{aA}\^B J_{bB}\^C   &=& -\ft14 \eta_{ab}\delta_A\^C
                        +\ft12 \eps_{abc}J_{cA}\^C, \zl
J\cc_{aA}\^B J\cc_{bB}\^C   &=& -\ft14 \eta_{ab}\delta_A\^C
                        +\ft12 \eps_{abc}J\cc_{cA}\^C, \zl
J_{aA}\^B J\cc_{bB}\^C &=& J\cc_{bA}\^B J_{aB}\^C.
\eeq
Self-duality:
\beq[self-dual]
\eps_{AB}\^{CD}J_{aCD} = 2\i J_{aAB}, \qquad
\eps_{AB}\^{CD}J\cc_{aCD} = - 2\i J\cc_{aAB}.
\eeq
Other useful formulae:
\beq[other]
\eps_{abc} J_{aAB} J_{bCD} &=&
    \ft12 \big(  \eta_{AD} J_{cBC} +  \eta_{BC} J_{cAD}
                - \eta_{AC} J_{cBD} -  \eta_{BD} J_{cAC}  \big), \zl
J_{aAB} J_{aCD} &=& \ft12 \eta_{A[C} \eta_{D]B}
                    - \ft\i4 \eps_{ABCD}.
\eeq

\end{document}